\documentclass[superscriptaddress,groupedaddress,nofootnoteinbib,12pt]{article} 
\usepackage{graphicx}
\usepackage{color}
\usepackage{dcolumn}   
\usepackage{bm}     
\usepackage{bbm}       
\usepackage{amssymb}  
\usepackage{amsmath}
\usepackage{sectsty}
\usepackage{latexsym}
\usepackage{float}
\usepackage{ifthen}
\usepackage{caption,subfig}
\usepackage{enumerate}
\usepackage{url}
\usepackage{caption,subfig}
\usepackage{feynmf}	
\usepackage{jcappub}
\usepackage{amsopn}

\usepackage{amsfonts}
\usepackage{multirow}
\usepackage{array}
\usepackage{booktabs}
\usepackage{rotating}

\def\clap#1{\hbox to 0pt{\hss#1\hss}}

\def\({\left(}
\def\){\right)}
\def\[{\left[}
\def\]{\right]}
\def\bea{\begin{eqnarray}}
\def\eea{\end{eqnarray}}
\def\be{\begin{equation}}
\def\ee{\end{equation}}
\def\ba{\begin{eqnarray}}
\def\ea{\end{eqnarray}}
\def\beq{\begin{eqnarray}}
\def\eeq{\end{eqnarray}}
\def\mpl{M_{\rm Pl}}
\def\d{\mathrm{d}}

\newcommand{\lambdat}{\tilde{\lambda}}

\newcommand{\LL}{\mathcal{L}}
\newcommand{\Tr}{{\rm Tr}}
\newcommand{\Mm}{\hat{M}}
\newcommand{\Wm}{\hat{W}}
\newcommand{\Rm}{\hat{R}}
\newcommand{\gm}{\hat{g}}
\newcommand{\Pm}{\hat{P}}
\newcommand{\Tm}{\hat{T}}
\newcommand{\Omegam}{\hat{\Omega}}
\newcommand{\rhot}{\tilde{\rho}}
\newcommand{\pt}{\tilde{p}}

\newcommand{\Id}{\mathbbm 1}

\def\clap#1{\hbox to 0pt{\hss#1\hss}}

\renewcommand{\geq}{\geqslant}
\renewcommand{\leq}{\leqslant}

\newcommand{\mS}{\mathcal{S}}
\newcommand{\Od}{\mathcal{O}}

\def\P{\hat{P}}

\definecolor{forestgreen}{rgb}{0.133,0.545,0.133}
\newboolean{editorial}
\setboolean{editorial}{true}
\newcommand{\editorial}[2]{\ifthenelse{\boolean{editorial}}{\textcolor{red}{[\textsf{\textbf{{#1}}}: }\textcolor{blue}{\textsf{{#2}}}\textcolor{red}{]}}{}}

\renewcommand{\d}{\mathrm{d}}
\renewcommand{\vec}[1]{\bm{\mathrm{{#1}}}}

 \def\be   {\begin{equation}}   \def\ee   {\end{equation}}

 \def\ba  {\begin{eqnarray}}   \def\ea  {\end{eqnarray}}

\renewcommand{\thesubsection}{\arabic{section}.\arabic{subsection}}

\begin{document}


\title{Infrared lessons for ultraviolet gravity: the case of massive gravity and Born-Infeld}

\author{Jose Beltr\'an Jim\'enez$^a$, Lavinia Heisenberg$^{b,c}$, Gonzalo J.  Olmo$^{d,e}$}
\affiliation{$^a$Centre for Cosmology, Particle Physics and Phenomenology,
Institute of Mathematics and Physics, Louvain University,
2 Chemin du Cyclotron, 1348 Louvain-la-Neuve, Belgium}
\affiliation{$^{b}$Perimeter Institute for Theoretical Physics, 
31 Caroline St. N,\\ Waterloo, Ontario, Canada N2L 2Y5
}
\affiliation{$^{c}$D\'epartement de Physique  Th\'eorique and Center for Astroparticle Physics,\\
Universit\'e de Gen\`eve, 24 Quai E. Ansermet, CH-1211  Gen\`eve, Switzerland}
\affiliation{$^{d}$Depto. de F\'{i}sica Te\'{o}rica \& IFIC, Universidad de Valencia - CSIC,\\ 
Calle Dr. Moliner 50, Burjassot 46100, Valencia, Spain}
\affiliation{$^{e}$Depto. de F\'isica, Universidade Federal da Para\'\i ba,  Cidade Universit\'aria, s/n - Castelo Branco, 58051-900 Jo\~ao Pessoa, Para\'\i ba, Brazil}

	\emailAdd{jose.beltran@uclouvain.be}
	\emailAdd{Lavinia.Heisenberg@unige.ch}
	\emailAdd{gonzalo.olmo@csic.es}
	
\abstract{We generalize the ultraviolet sector of gravitation via a Born-Infeld action using lessons from massive gravity. The theory contains all of the elementary symmetric polynomials  and is treated in the Palatini formalism. We show how the connection can be solved algebraically to be the Levi-Civita connection of an effective metric. The non-linearity of the algebraic equations yields several branches, one of which always reduces to General Relativity at low curvatures. We explore in detail a  {\it minimal} version of the theory, for which we study solutions in the presence of a perfect fluid with special attention to the cosmological evolution. In vacuum we recover  Ricci-flat solutions, but also an additional physical solution corresponding to an Einstein space. The existence of two physical branches remains for non-vacuum solutions and, in addition, the branch that connects to the Einstein space in vacuum is not very sensitive to the specific value of the energy density. For the branch that connects to the General Relativity limit we generically find three behaviours for the Hubble function depending on the equation of state of the fluid, namely: either there is a maximum value for the energy density that connects continuously with vacuum, or the energy density can be arbitrarily large but the Hubble function saturates and remains constant  at high energy densities, or the energy density is unbounded and the Hubble function grows faster than in General Relativity. The second case is particularly interesting because it  could offer an interesting inflationary epoch even in the presence of a dust component. Finally, we discuss the possibility of avoiding certain types of singularities within the minimal model.
}

\maketitle

\section{Introduction}
\label{sec:intro}

Over the last century, physicists have devoted much effort  to determine and understand the content, structure and evolution of the Universe. In this task, General Relativity (GR) has played a fundamental role in shaping the current standard model of cosmology. However, this theory faces a number of challenges from both theoretical and observational perspectives. From the theoretical side, the existence of cosmological and black hole singularities are disturbing issues that suggest the potential unsuitability of the theory in its standard form for the description of gravitational phenomena at sufficiently high energies. A consistent combination of GR with quantum field theory in a quantum theory of gravity is another fundamental open question, although GR is perfectly valid as an effective field theory  for most physical situations. At Ultra-Violet (UV) scales, the cosmological constant problem represents a challenging puzzle which reflects the large discrepancy between observations and the theoretical predictions using standard techniques of quantum field theory \cite{Weinberg:1988cp,Martin:2012bt}. On the other hand, the observational evidence supporting the recent accelerated expansion of the Universe  could signal a failure of GR at the largest distance scales and, thus, motivates Infra-Red (IR) modifications of the gravitational interaction. Additionally, the need for some kind of dark matter to account for certain astrophysical and cosmological observations may also be an indication of the need to go beyond GR.  All these facts have triggered a burst of activity searching for consistent modifications of gravity in both the UV and the IR domains of the theory.\\

The existence of cosmological and black hole singularities, scenarios in which curvature scalars may grow without bound, is typically attributed to the classical nature of GR. For this reason, it is traditionally argued that quantum gravitational effects should take care of those pathologies at sufficiently high energies or at scales of order the Planck length. Almost by definition, it seems generally accepted that a quantum theory of gravity will regularize curvature divergences. However, one might dare to consider a different situation in which getting rid of curvature singularities is not included among the important roles of quantum gravity. In analogy with the transition from non-relativistic to relativistic mechanics, where the velocity of a body is bounded by mechanisms not involving quantum physics, one might consider a scenario in which classical gravitation is somehow improved in such a way that curvature scalars are generically regular. The quantization of gravity in such a regular scenario could thus proceed in a way substantially different from the current approaches. \\

A class of gravity models recently considered in the literature makes contact with the above ideas and, at the same time, also establishes interesting connections with extensions of gravity that affect the infrared sector. Originally proposed by Deser and Gibbons \cite{Deser:1998rj}, a Born-Infeld-like gravity Lagrangian has been studied recently with interesting applications in cosmology and black hole scenarios \cite{Ketov:2001dq,Wohlfarth:2003ss,Comelli:2004qr,Banados:2008rm,Avelino:2012ge,Pani:2012qb,Avelino:2012ue,Olmo:2013gqa,Yang:2013hsa,Bouhmadi-Lopez:2014jfa,Gullu:2010pc,Gullu:2010wb,Gullu:2010st,Gullu:2010em}. Born-Infeld gravity represents a specific infinite order higher curvature modification of GR. Extensions  with higher order curvature terms usually contain ghostly degrees of freedom and/or are not unitary even though they can be renormalizable. Nevertheless, when formulated \`{a} la Palatini, these theories of gravity are ghost-free and may avoid cosmological and black hole singularities in some cases. Though the theory does not cure all curvature divergences, it does provide interesting new insights along the line proposed above aimed at building a classical theory of gravity with bounded curvature scalars. The theory in the matrix representation is characterized by the determinant of a matrix defined as $\sqrt{-\det{(\hat{g}+\lambda^{-2} \hat{R}(\Gamma)})}$ where $\hat{g}$ is the space-time metric, $\hat{R}$ is the Ricci tensor and $\lambda$ is an energy scale. This gravity Lagrangian is similar to the original Born-Infeld electromagnetic Lagrangian \cite{Ketov:2001dq}, $\sqrt{-\det( \hat{g}+ \lambda^{-2} \hat{F} )} $, where $ \hat{F}$ is the electromagnetic field strength,  proposed to bound the self-energy and field strength of the electron. The above gravity action recovers GR in the limit $\lambda\to \infty$ in the same way as the Born-Infeld electromagnetic Lagrangian recovers Maxwell's theory when $\lambda\to \infty$. Note that the bound on the speed of material particles can be achieved by promoting the non-relativistic free Lagrangian $-\frac{1}{2}m v^2$ to the Born-Infeld-like Lagrangian $-mc^2 \sqrt{1-v^2/c^2}$, which recovers the non-relativistic free Lagrangian in the limit $m c^2\to \infty$ \cite{Ferraro:2009zk}. The Born-Infeld gravity \`{a} la Palatini can thus be seen as an attempt to build a dynamical classical geometry with bound curvature scalars via the Born-Infeld algorithm. We note that Born-Infeld like actions seem to be very fundamental, as they naturally appear in the description of the electromagnetic field of certain $D$-branes \cite{Gibbons:2001gy} 
.\\ 

Following \cite{Odintsov:2014yaa}, the Born-Infeld gravity lagrangian can also be written as $ \sqrt{-g}\sqrt{\det(\hat\Omega)}$, where $\hat\Omega=\hat g^{-1} \hat q$ and $\hat q$ denotes $\hat{q}\equiv \hat{g}+\lambda^{-2}\hat{R}$. This notation puts forward that  the matrix $\hat \Omega$ plays a fundamental role in the definition of the theory. Interestingly, a similar object has also been identified recently in a different context as a basic building block of a consistent, ghost-free massive theory of the graviton \cite{deRham:2010kj}. In that scenario, the potential terms that give mass to the graviton are defined in terms of an object $\hat M=\sqrt{\hat g^{-1} \hat f}$, where $\hat f$ represents a reference metric, which can be dynamical in some models. In this work we will use some results obtained within the framework of massive gravity theories to reinterpret the gravity action of Born-Infeld and  extend it to a larger family of theories. \\

 Massive gravity is one important class of infrared modifications of gravity  in which the mediator of gravitational interactions has a small mass. The construction of a well-defined, non-linear theory of massive gravity without any pathologies like the Boulware-Deser ghost has been a challenging problem over the last decades. However, in recent years such a theory has been successfully constructed by de Rham, Gabadadze and Tolley \cite{deRham:2010ik,deRham:2010kj}. This theory generalizes the linear Fierz-Pauli action without propagating ghost degrees of freedom and contains only second order equations of motion. Its successful  realization relies on a very specific structure of a 2-parameter family interactions. Very soon after that Hassan and Rosen have realized that this very specific structure of the interactions was automatically fulfilled by writing the interactions in terms of a deformed determinant  $\det{(\Id+\beta_n\hat K)}$ \cite{Hassan:2011vm}, where the matrix $\hat K=\sqrt{\hat g^{-1}\hat f}-\Id$ encodes the potential interactions of the graviton with the square root structure. This deformed determinant on the other hand can be written in terms of the symmetric polynomials of the matrix $\hat K$. The determinant plays a very crucial role. An equivalent description of the interactions in terms of the Stueckelberg fields requires that the scalar Stueckelberg field $\pi$ in zeroth order of metric fluctuations $h_{\mu\nu}=0$ must appear in the action only in total derivative terms. Exactly the antisymmetry property of the Levi-Civita tensors in the determinant $\det{(\delta^{\mu}_{\nu}+\partial^\mu\partial_\nu\pi)}$ guarantees that. Furthermore, the theory of massive gravity can be promoted to a theory of bigravity by invoking explicit dynamics for the reference metric $f$ \cite{Hassan:2011zd}. The ghost-free interactions between the two metrics are still given by the determinant  $\det{(\Id+\hat K)}$ with the difference that there is also an additional kinetic term for the $f$ metric. There is nothing special about one metric versus the other since the theory is symmetric under the exchange of the two metrics. Therefore naively one could attempt to couple both metrics to external matter fields even though there might be restrictions due to the existence of ghost degrees of freedom \cite{Akrami:2013ffa,Akrami:2014lja,Yamashita:2014fga,deRham:2014naa,Noller:2014sta,deRham:2014fha,Gumrukcuoglu:2014xba,Heisenberg:2014rka}. Another crucial point is that massive gravity in a given limit reproduces a given important class of scalar-tensor theories: the Galileon theory \cite{Nicolis:2008in}. The Galileon interactions represent an interesting subclass of Horndeski interactions \cite{Horndeski:1974wa} with shift and Galileon symmetry and they have only second order equations of motion. The decoupling limit of massive gravity contains these interactions automatically \cite{deRham:2010ik, deRham:2010tw}. Exactly as in massive gravity the Galileon interactions can also be written in terms of a deformed determinant $\det{(\delta^{\mu}_{\nu}+a\partial^\mu\partial_\nu\pi+b\partial^\mu\pi\partial_\nu\pi )}$ \cite{Tasinato:2013wna}. As can be seen, in all these theories, the interactions appear in a very specific form of a deformed determinant, to which we will pay a specific attention when we construct our generalization of the Born-Infeld gravity theory. In this work we will combine the Born-Infeld gravity with the ideas of massive gravity and extend the Born-Infeld action to include all of the symmetric polynomials.\\

By comparing the structures of the basic building blocks of Born-Infeld gravity and the potential term in massive gravity, the Born-Infeld Lagrangian can be written either as $\sqrt{-g}\sqrt{\det\hat\Omega}$ \cite{Odintsov:2014yaa} or as $\sqrt{-g}\det(\sqrt{\hat\Omega})$. The identification of the matrix $\sqrt{\hat\Omega}$ as the basic element of the action and the fact that $\det\sqrt{\hat\Omega}$ is nothing but the highest order elementary symmetric invariant polynomial of this matrix motivates us to propose an extension involving other invariant polynomials of $\sqrt{\hat\Omega}$. After explaining in detail this construction \ref{sec:theory}, we derive the field equations of the resulting theory \ref{sec_EOM}. Then we concentrate in more detail on the specific case of the minimal extension of the theory, which consists only of the first polynomial \ref{min_BI}, and consider different applications aimed at exploring the behavior of these new theories in scenarios with cosmological singularities \ref{cosmology_min_BI}.   \\

Throughout the paper, we use the metric signature convention $(-,+,+,+)$ and units in which the speed
of light and the Planck constant are unity $c=\hbar=1$, and the reduced Planck mass is $M_{\rm Pl}^2=1/\sqrt{8\pi G}$. We follow the convention  $R_{\alpha\beta}=\partial_\mu \Gamma^\mu_{\beta\alpha}-\partial_\beta \Gamma^\mu_{\mu\alpha}+\Gamma^\mu_{\mu\nu}\Gamma^\nu_{\beta\alpha}-\Gamma^\mu_{\beta\nu}\Gamma^\nu_{\mu\alpha}$ for the Ricci tensor  and $\mathcal{T}^{\alpha}_{\mu\nu}=\Gamma^\alpha_{[\mu\nu]}$ for the torsion. Some contractions of rank-2 tensors are denoted by
${M}^{\mu}_{~\mu}=[{ M}]=\Tr{(M)}$,~
${ M}^{\mu}_{~\nu}{M}^{\nu}_{~\mu}=[{ M}^2]=\Tr{(M^2)}$,~
${ M}^{\mu}_{~\alpha}{M}^{\alpha}_{~\beta}{M}^{\beta}_{~\mu}=[{ M}^3]$, 
and so on.


\section{The theory}\label{sec:theory}

\subsection{Born-Infeld electromagnetism}
Motivated by finding a regularization scheme for the singularities that appear in classical electromagnetism at small scales (UV regime), Born and Infeld suggested to construct a nonlinear extension of the Maxwell action in the form \cite{Born:1934gh}
\begin{equation}
\mathcal S_{\rm BIE}=-\lambda^4\int \d^4x \left[\sqrt{-\det{(\eta_{\mu\nu}+\lambda^{-2} F_{\mu\nu})}}-1\right],
\end{equation}
where $\eta_{\mu\nu}$ is the Minkowski metric, $F_{\mu\nu}=\partial_\mu A_\nu-\partial_\nu A_\mu$ the strength tensor of the electromagnetic potential $A_\mu$ and $\lambda$ some energy scale. The form of this action follows a close analogy to that of a relativistic point particle $S_{\rm pp}=-m c^2\int\d t\sqrt{1-v^2/c^2}$. Very much like in special relativity the speed of light serves as an upper limit for speeds, in Born-Infeld electromagnetism the scale $\lambda$ gives the maximum allowed value for electromagnetic fields. The above action can  be alternatively written as
\bea
\mathcal S_{\rm BIE}&=&-\lambda^4\int \d^4x \left[\sqrt{1+\frac{1}{2\lambda^4}F_{\mu\nu}F^{\mu\nu}-\frac{1}{16\lambda^8}(F_{\mu\nu}\tilde{F}^{\mu\nu})^2}-1\right]\\\nonumber
&=&-\lambda^4\int \d^4x \left[\sqrt{1-\frac{\vec{E}^2-\vec{B}^2}{\lambda^4}-\frac{(\vec{E}\cdot\vec{B})^2}{\lambda^8}}-1\right]
\eea
with $\tilde{F}^{\mu\nu}\equiv\frac12\varepsilon^{\mu\nu\alpha\beta} F_{\alpha\beta}$ the Hodge dual of $F_{\alpha\beta}$ and in the second line we have used the usual definitions for the electric and magnetic fields $E_i=-F_{0i}$ and $B_i=\frac12 \epsilon_{ink}F^{jk}$. For small electromagnetic fields $\vert F_{\mu\nu}\vert\ll \lambda^2$, the leading order contribution to the action reads
\bea
\mathcal S_{\rm BIE}(\vert F_{\mu\nu}\vert\ll \lambda^2)&\simeq&-\frac14\int \d^4x F_{\mu\nu}F^{\mu\nu}
\eea
and we see that the theory reduces to usual Maxwell electrodynamics. However, for fields of order $\lambda^2$ the self-interaction terms become important and prevent the unlimited growth of electric and magnetic fields so that UV singularities are regularized. In particular, the self-energy of point-like charges becomes finite. One important feature of Born-Infeld electromagnetism is that it preserves the hyperbolicity of the field equations, thus, guaranteeing causal propagation.

\subsection{Born-Infeld inspired gravity}

Inspired by  Born-Infeld electrodynamics, Deser and Gibbons \cite{Deser:1998rj} suggested a similar construction for gravity. They started by considering an action of the form

\begin{equation}
\mathcal S_{\rm BIG}=\lambda^4\int \d^4x  \sqrt{-\det{(g_{\mu\nu}+\lambda^{-2} R_{\mu\nu}+c_1 X_{\mu\nu})}} 
\label{eq:SBIG1}
\end{equation}
where $g_{\mu\nu}$ is the spacetime metric tensor, $R_{\mu\nu}$ is the metric Ricci tensor and $X_{\mu\nu}$ contains terms of quadratic and higher orders in $R_{\mu\nu}$ and must be chosen to remove ghosts from the spectrum of the theory. The scale $\lambda$ must be tuned to recover GR at low curvatures.

It was argued in \cite{Deser:1998rj} that, unlike in the original Born-Infeld construction for electromagnetism, there was no clear criterion to make an analogous construction for gravity. The ghost problem for quadratic and higher order terms that required the introduction of $X_{\mu\nu}$ in (\ref{eq:SBIG1}) arises in the pure metric formalism of the theory, i.e., when the Ricci tensor is defined in terms of the symmetric and metric compatible Levi-Civita connection. However, if we treat the connection as an independent object, i.e., in the Palatini formulation of the theory, ghosts do not arise anymore from higher order terms in $R_{\mu\nu}$. Thus, we could give up the metric formulation of the theory and use a metric-affine variational principle instead. Such formulation resembles Eddington's theory, which is a pure affine theory described by the action
\begin{equation}
\mathcal S_{\rm Ed}=\lambda^4\int \d^4x \sqrt{\det{ R_{(\mu\nu)}(\Gamma)}} 
\end{equation}
where $R_{(\mu\nu)}$ is the symmetric part of the Ricci tensor and the connection $\Gamma^\alpha_{\mu\nu}$ (that is assumed to be symmetric, i.e., no torsion is present) appears as the fundamental geometric quantity. In the spirit of Eddington's theory, it is natural to consider the {\it simplified} version of the Born-Infeld inspired action (\ref{eq:SBIG1}) in the Palatini formalism 
\begin{equation}
\mathcal S_{\rm BIP}=\lambda^4\int \d^4x \left[\sqrt{-\det{(g_{\mu\nu}+\lambda^{-2} R_{\mu\nu}(\Gamma)})}- \sqrt{-\det({g_{\mu\nu})}}\right]
\end{equation}
where the second term is introduced to have Minkowski vacuum solutions, as we will see below. One can write this action also as
\begin{equation}
\mathcal S_{\rm BIP}=\lambda^4\int \d^4x\sqrt{-g} \left[\sqrt{\det{(\delta^\mu{}_\nu+\lambda^{-2}P^\mu{}_\nu)}}-1\right]
\end{equation}
with $P^\mu{}_\nu=g^{\mu\alpha}R_{\alpha\nu}(\Gamma)$. Now, by commuting the determinant and the square root\footnote{This is always an admissible operation provided the square root exists. The square root of a matrix is defined by $\sqrt{M}\sqrt{M}=M$ so, taking  determinants in both sides we obtain $(\det\sqrt{M})^2=\det M$ from which it trivially follows that $\det \sqrt{M}=\sqrt{\det M}$.}, we can write the action in a more suggestive form
\begin{equation}
\mathcal S_{\rm BIP}=\lambda^4\int \d^4x\sqrt{-g} \left[\det\sqrt{1+\lambda^{-2}\P}-1\right]
\end{equation}
where we have denoted by $\P$ the corresponding matrix. Note that the expression under the square root can also be expressed as
 \begin{equation}
(1+\lambda^{-2} \hat{P})^\mu{}_\nu=\delta^\mu_\nu+\lambda^{-2}g^{\mu\alpha}R_{\alpha\nu}(\Gamma)=g^{\mu\alpha}(g_{\alpha\nu}+\lambda^{-2} R_{\alpha\nu}(\Gamma))=g^{\mu\alpha}q_{\alpha\nu}=(\hat{g}^{-1}\hat{q})^\mu{}_\nu
\end{equation}
where we defined $q_{\alpha\nu}\equiv g_{\alpha\nu}+\lambda^{-2} R_{\alpha\nu}(\Gamma)$. Therefore, the Born-Infeld inspired action in the Palatini formalism can simply be written as \cite{Odintsov:2014yaa}
\begin{equation}
\mathcal S_{\rm BIP}=\lambda^4\int \d^4x\sqrt{-g} \left(\det{(\sqrt{\hat{g}^{-1}\hat{q}})}-1\right)\,.
\end{equation}
After defining the matrix $\hat{M}\equiv \sqrt{\hat{g}^{-1}\hat{q}}=\sqrt{1+\lambda^{-2}\hat{P}}$ we can simply write the action as
\begin{equation}
\mathcal S_{\rm BIP}=\lambda^4\int \d^4x\sqrt{-g} \left(\det\hat{{M}}-1\right).
\label{BIP2action}
\end{equation}
This is the form of the Born-Infeld gravity action that will serve as our starting point to motivate our extension of this theory.

\subsection{Generalizing Born-Infeld inspired gravity theories}
Now we are ready to introduce our generalization of Born-Infeld gravity (for other recent extensions see \cite{Makarenko:2014lxa,Makarenko:2014nca,Odintsov:2014yaa}). It is a trivial observation that action \ref{BIP2action} contains two of the elementary invariant symmetric polynomials of the matrix $\hat{M}$. The extension we propose and will explore in this work is quite suggestive. Instead of only considering these two symmetric polynomials of the matrix $\hat{M}$, we will generalize the action to contain all of the remaining polynomials. Our generalized Born-Infeld action \`a la Palatini is thus given by \footnote{The cosmological constant term is now included in $e_0(\hat{M})$.}
\begin{equation}\label{gen_Born_Infeld}
\mathcal S_{GBI}=\lambdat^4\int \d^4x\sqrt{-g}\sum_{n=0}^4 \beta_ne_n(\hat{M}).
\end{equation}
In this action we have introduced an additional scale $\lambdat$ that will be fixed below by the requirement to recover GR at low curvatures, $\beta_n$ are free dimensionless parameters and $e_n(\Mm)$ are the elementary symmetric polynomials defined as
\begin{eqnarray}\label{eq:polynomials}
e_0(\Mm) &=& 1,\nonumber\\
e_1(\Mm) &=& [\Mm] ,\nonumber\\
e_2(\Mm) &=& \frac{1}{2!}\Big([\Mm]^2- [\Mm^2]\Big),  \nonumber\\
e_3(\Mm) &=&\frac{1}{3!}\Big( [\Mm]^3- 3[\Mm][\Mm^2]+2[\Mm^3] \Big),\nonumber\\
e_4(\Mm) &=&\frac{1}{4!}\Big([\Mm]^4-6[\Mm]^2[\Mm^2]+8[\Mm][\Mm^3]+3[\Mm^2]^2-6[\Mm^4]   \Big).
\end{eqnarray}
The fourth symmetric polynomial is nothing but the determinant $e_4(\hat{M})=\det\hat {M}$ and is precisely the one appearing in the original Born-Infeld inspired gravity theory. These symmetric polynomials are invariant under any transformation $\hat{R}^{-1}\hat{M}\hat{R}$ whose inverse exist $\hat{R}^{-1}\hat{R}=\Id$. We can also write them in terms of the Levi-Civita tensors
\begin{eqnarray}
e_0 &=&-\frac{1}{4!} \varepsilon_{\mu\nu\alpha\beta}\varepsilon^{\mu\nu\alpha\beta}  \nonumber\\
e_1 &=&-\frac{1}{3!} \varepsilon_{\mu\nu\alpha\beta}\varepsilon^{\rho\nu\alpha\beta}M^{\mu}_{\;\;\rho} \nonumber\\
e_2 &=&-\frac{1}{2!2!} \varepsilon_{\mu\nu\alpha\beta}\varepsilon^{\rho\sigma\alpha\beta}M^{\mu}_{\;\;\rho}M^{\nu}_{\;\;\sigma}\nonumber\\
e_3 &=&- \frac{1}{3!}\varepsilon_{\mu\nu\alpha\beta}\varepsilon^{\rho\sigma\delta\beta}M^{\mu}_{\;\;\rho}M^{\nu}_{\;\;\sigma}M^{\alpha}_{\;\;\delta} \nonumber\\
e_4 &=&-\frac{1}{4!} \varepsilon_{\mu\nu\alpha\beta}\varepsilon^{\rho\sigma\delta\gamma}M^{\mu}_{\;\;\rho}M^{\nu}_{\;\;\sigma}M^{\alpha}_{\;\;\delta}M^{\beta}_{\;\;\gamma} \nonumber\\
\end{eqnarray}
with $M^\mu{}_\nu=\sqrt{\delta^\mu_\nu+\lambda^{-2}g^{\mu\alpha}R_{\alpha\nu}(\Gamma)}=\sqrt{g^{\mu\alpha} q_{\alpha\nu}}$. \\

As commented above, one of the motivations to consider this extension of the Born-Infeld inspired action comes from the realization of ghost-free theories of massive (bi-)gravity. Indeed, there one has non-trivial interactions of a similar form \cite{deRham:2010kj, Hassan:2011vm}
\begin{equation}
\mathcal S_{MG}=\int d^4x\sqrt{-g} \sum_{n=0}^4 \frac{\beta_n}{n!(4-n)!} e_n(\sqrt{g^{-1}f})
\end{equation}
where $f_{\mu\nu}$ is the reference metric, which is also dynamical in the case of bigravity. The square root structure of the interactions in massive gravity plays a very crucial role: it guarantees the absence of ghostly degrees of freedom. This striking similarity motivates us to consider the full set of invariant symmetric polynomials also in the case of the Born-Infeld inspired theories. Actually, not only in massive (bi-)gravity, but also in some given interesting subgroups of Horndeski interactions, like the Galileon interactions, we find this form of a deformed determinant $\det{(\delta_{\mu\nu}+a\partial_\mu\partial_\nu\pi+b\partial_\mu\pi\partial_\nu\pi )}$. \\

\subsubsection{Low curvature limit}
As it happens with the original Born-Infeld gravity theory, the low curvature limit of our generalized version successfully recovers General Relativity. This can be easily checked by expanding the elementary symmetric polynomials to the corresponding IR limit, namely $\vert g^{\mu\alpha}R_{\alpha\nu}\vert\ll\lambda^2$. In such a limit, we have that $\hat{M}^n\simeq \Id+\frac n2\lambda^{-2}\hat{P}$ (with $\hat{P}=\hat{g}^{-1}\hat{R}(\Gamma)$) and the action becomes:
\begin{equation}\label{genPal}
\mathcal S\simeq\lambdat^4\int \d^4x\sqrt{-g}\left[ \left(\beta_0+4\beta_1+6\beta_2+4\beta_3+\beta_4  \right)+\frac{1}{2\lambda^2}\left(\beta_1+3\beta_2+3\beta_3+\beta_4  \right) g^{\mu\nu}R_{\mu\nu}(\Gamma)  \right] 
\end{equation}
which coincides with the Einstein-Hilbert action in the Palatini formalism supplemented with a cosmological constant term (which can be cancelled by tuning the parameter $\beta_0$). As it is well-known, the Palatini formulation of the Einstein-Hilbert action is not exactly GR because there is a projective invariance with $\Gamma^\alpha_{\mu\nu}\rightarrow\Gamma^\alpha_{\mu\nu}+\delta^\alpha_\nu\xi_\mu$ for an arbitrary $\xi_\mu$. This symmetry of the Einstein-Hilbert action leaves four components of the connection undetermined that must be fixed by additional conditions, like the vanishing of the torsion vector or the trace of the non-metricity tensor. This can actually be implemented in the action itself by the inclusion of suitable lagrange multipliers. In the case of the pure Einstein-Hilbert action, this symmetry is exact, but in our generalized Born-Infeld gravity (and also in its original version), the projective invariance only appears as an approximate low curvature  symmetry that will be broken by higher order contributions and, thus, the necessary conditions to recover GR must be imposed by hand.

Notice that to recover GR we need to identify 
\be
\frac{\lambdat^4}{\lambda^2}\left(\beta_1+3\beta_2+3\beta_3+\beta_4  \right)= \mpl^2. 
\ee
This fixes the scale $\lambdat$ and leaves $\lambda$ as the only new dimensionful parameter in the theory which controls the scale at which high curvature effects become important (assuming that $\beta_n$ are of order 1). Finally, we should mention that what we have shown is the existence of one branch of solutions that will reduce to GR in the limit of low curvatures. However, in the full theory it is expected to find several branches of solutions and some of them will not reduce to GR in such a limit. We will come back to this point later in specific examples.

\subsubsection{High curvature limit}

We have seen that the low curvature limit of the extended Born-Infeld theories of gravity reduce to the Einstein-Hilbert action in the Palatini formalism. Now, let us consider the high curvature limit where $\vert g^{\mu\alpha}R_{\alpha\nu}\vert\gg\lambda^2$. For that, let us first note that in such a limit we have $\hat{M}\simeq\sqrt{\lambda^{-2} \hat{g}^{-1}\hat{R}}$ so that the $n$-th polynomial will be $e_n\sim {\mathcal O}(\hat{M}^n)\simeq{\mathcal O}(\lambda^{-n}(\hat{g}^{-1}\hat{R})^{n/2})$. If we have all the polynomials, then the $4^{\rm th}$ one will dominate and we will recover a type of Eddington action
\be
\mS\simeq\beta_4\frac{\lambdat^4}{\lambda^4}\int\d^4x\sqrt{\det R_{\mu\nu}(\Gamma)}.
\ee
However, whereas in the original Eddington theory only the symmetric part of the Ricci tensor is taken, here the whole Ricci curvature contributes, including its antisymmetric part. 

On the other hand, if we consider the case where only the polynomials up to $k$-th order  contribute to the action, that one will be the relevant term for the high curvature limit, i.e., only the highest polynomial will contribute. The high curvature limit for all the polynomials is given in the following:
\begin{eqnarray}\label{eq:polynomialsHEL}
e_0 &\simeq& 1\nonumber\\
e_1 &\simeq&\vert\lambda\vert^{-1} \left[\sqrt{\Pm}\;\right] \nonumber\\
e_2 &\simeq& \frac{\lambda^{-2}}{2!}\left(\left[\sqrt{\Pm}\;\right]^2- [\Pm]\right)  \nonumber\\
e_3 &\simeq&\frac{\vert\lambda\vert^{-3}}{3!}\left( \left[\sqrt{\Pm}\;\right]^3- 3\left[\sqrt{\Pm}\;\right][\Pm]+2\left[\Pm^{3/2}\right] \right)\nonumber\\
e_4 &\simeq&\frac{\lambda^{-4}}{4!}\left(\left[\sqrt{\Pm}\;\right]^4-6\left[\sqrt{\Pm}\;\right]^2[\Pm]+8\left[\sqrt{\Pm}\;\right]\left[\Pm^{3/2}\right] +3[\Pm]^2-6[\Pm^2]   \right)
\end{eqnarray}
where we remind that $P^\mu{}_\nu=g^{\mu\alpha}R_{\alpha\nu}(\Gamma)$. We find particularly interesting  the case of $e_2$ since it leads to a suggestive modification of the Einstein-Hilbert action without new dimensionfull parameters. If we forget for a moment about its origin as the high curvature limit of our generalized Born-Infeld theory and take it as the starting action it reads
\be
\mS_{\rm EEH}=\tilde{m}^2\int \d^4x\sqrt{-g} \left( \Big[\hat{g}^{-1}\hat{R}\Big]-\Big[\sqrt{\hat{g}^{-1}\hat{R}}\Big]^2 \right)
\ee
with $\tilde{m}$ some scale. This theory could even be treated in the metric formalism. Now if we interpret the operation of tracing as a type of averaging, the above action can be interpreted as being the variance of $\sqrt{\hat{g}^{-1}\hat{R}}$. Despite its amusing interpretation, its physical viability is dubious since it likely leads to observational conflicts with local gravity tests and the hyperbolicity of the field equations might also be violated due to the square root structure with differential operators. However, these  issues should be explored before reaching a definite conclusion.


\section{Equations of motion}\label{sec_EOM}
In the following we will compute the covariant equations of motion for our generalized Born-Infeld inspired gravity theory. To make the derivation more transparent, let us start with 
a Lagrangian depending on one single polynomial $e_n(\hat M)$, whose action is
\begin{eqnarray}
 \mS_{n}=\lambdat^4\int \d^4x\sqrt{-g}\,  e_n(\hat M) \ .  
\end{eqnarray}
Variation of this action leads to 
\begin{eqnarray}\label{eq:var_0}
 \delta \mS_{n}=\lambdat^4\int \d^4x\sqrt{-g} \left[ -\frac{1}{2} g_{\alpha\beta}\delta g^{\alpha\beta}e_n(\hat M)+\delta e_n(\Mm)\right] \,, 
\end{eqnarray}
where we have used $\delta\sqrt{-g}=-\frac12\sqrt{-g}g_{\alpha\beta}\delta g^{\alpha\beta}$. The variation $\delta e_n(\Mm)$ can be computed as follows. Since the polynomial $e_n(\hat M)$ is a function of the traces of powers of $\hat M$, we can write (recall that $[\Mm^k]\equiv \Tr(\hat M^k)$) 
\begin{equation}
\delta e_n= \sum_{k=1}^nE^k_n \delta [M^k] \ ,
\end{equation}
where we have defined $E^k_n\equiv \frac{\partial e_n}{\partial [M^k]}$. The sum over $k$ must start from $k=1$ for the expression to be meaningful. Notice that for $n=0$ we simply have $e_0=1$ which has vanishing variation.
Now, since $\hat {M}= (\hat \Omega)^{1/2}$ (where $\hat\Omega=\Id +\lambda^{-2}\hat{g}^{-1}\hat{R}$), we get that\footnote{This can be easily proven by varying $\Mm^k\Mm^k=\Omegam^k$ to obtain $\delta\Mm^k\Mm^k+\Mm^k\delta\Mm^k=k\Mm^{2k-2}\delta\Omegam$. Then one multiplies by $\Mm^{-k}$ ant takes the trace to obtain the desired expression.}
\begin{equation}
\delta [M^k] = \frac{k}{2}\Tr[\hat M ^{k-2}\delta \hat \Omega].
\end{equation}
Now, since $\delta {\Omega^\alpha}_\beta =\lambda^{-2}(\delta g^{\alpha\gamma}R_{\gamma\beta}+g^{\alpha\gamma}\delta R_{\gamma\beta})$ we can rewrite the above variation as
\begin{equation}
\delta [M^k] = \frac{\lambda^{-2}k}{2}\left[R_{\gamma\beta} {(\hat M ^{k-2})^\beta}_\alpha\delta g^{\alpha\gamma}+{(\hat M ^{k-2})^\beta}_\alpha g^{\alpha\gamma}\delta R_{\gamma\beta} \right] 
\end{equation}
so that the action variation in (\ref{eq:var_0}) can be recast in the following form
\begin{eqnarray}\label{eq:var_1}
 \delta \mS_{n}&=&\lambdat^4\int \d^4x\sqrt{-g} \left[\left(
\sum_{k=1}^n\frac{\lambda^{-2}k  E^k_n}{4}\left(R_{\gamma\beta} {(\hat M ^{k-2})^\beta}_\alpha+R_{\alpha\beta} {(\hat M ^{k-2})^\beta}_\gamma\right) -\frac{1}{2} g_{\alpha\gamma}e_n(\hat M)\right)\delta g^{\alpha\gamma}\right. \nonumber \\ &+& \left. \sum_{k=1}^n\frac{\lambda^{-2}k  E^k_n}{2} {(\hat M ^{k-2})^\beta}_\alpha g^{\alpha\gamma}\delta R_{\gamma\beta} \right]  \ .  
\end{eqnarray}
The extension of this variation to the linear combination of polynomials of our action for the generalized Born-Infeld gravity now proceeds very easily since $\mS_{GBI}=\sum_{n =0}^4\beta_n\mS_n$. The variation of $\mS_{GBI}$ can thus be written as 
\begin{eqnarray}\label{eq:var_2}
 \delta \mS_{GBI}&=&\int\d^4x\sqrt{-g} \left[\left(\frac{\lambdat^4\lambda^{-2}}{4}
\left(R_{\alpha\lambda} {W^\lambda}_\beta+R_{\beta\lambda} {W^\lambda}_\alpha\right) -\frac{\LL_G}{2} g_{\alpha\beta}-\frac{1}{2}T_{\alpha\beta}\right)\delta g^{\alpha\beta}\nonumber \right.\\&+&\left. \frac{\lambdat^4\lambda^{-2}}{2}{ W^\alpha}_\lambda g^{\lambda\beta}\delta R_{\alpha\beta} \right]  \ ,  
\end{eqnarray}
where we have defined the total Lagrangian as $\LL_G\equiv \lambdat^4\sum_{n=0}^4  \beta_ne_n(\hat M)$, have added the stress-energy tensor $T_{\alpha\beta}$ coming from the variation of the matter sector, and have absorbed the sums over $k$ and $n$ in the matrix ${W^\lambda}_\beta$, which is defined as 
\begin{equation}
W^\lambda{}_\beta\equiv\sum_{n=1}^4 \beta_n\sum_{k=1}^n k  E_n^k (\Mm^{k-2})^\lambda{}_\beta \ .
\end{equation}
Notice that we have removed the term with $n=0$ from the sum over $n$ in the above definition because it corresponds to $e_0=1$, which does not actually depend on $\Mm$. We can let the sum over $k$ to run up to 4 since $E^k_n=0$ for $k>n$. Then, $E^k_n$ can be written as a square matrix whose components are given by 
\begin{equation}
E_n^k=\left( \begin{array}{cccc} 
                      e_0 & 0 & 0 & 0 \\
                      e_1 & -\frac{e_0}{2} & 0 & 0 \\
e_2 & -\frac{e_1}{2} & \frac{e_0}{3} & 0 \\
e_3 & -\frac{e_2}{2} & \frac{e_1}{3} &  -\frac{e_0}{4} \end{array}\right) \ ,
\end{equation}
where the index $k$ specifies the column and $n$ the row. The matrix $\hat W$ can thus be written as 
\begin{equation}
\hat W= f_1 \hat M^{-1}+f_2 \Id+f_3 \hat M+ f_4\hat M^2 \ , 
\end{equation}
where we have introduced the following definitions
\begin{eqnarray}
f_1&=& \beta_1 e_0+\beta_2 e_1+\beta_3 e_2+\beta_4 e_3\\
f_2&=& -(\beta_2 e_0+\beta_3 e_1+\beta_4 e_2)\\
f_3&=& \beta_3 e_0+\beta_4 e_1\\
f_4&=& -\beta_4 e_0\,.
\end{eqnarray} 
Since we are assuming the metric and connection have independent variations, the metric field equations that follow from (\ref{eq:var_2}) are 
\begin{equation}\label{eq:metric_var}
\frac{\lambdat^4\lambda^{-2}}{2}
\left(R_{\alpha\lambda} {W^\lambda}_\beta+R_{\beta\lambda} {W^\lambda}_\alpha\right) -\LL_G g_{\alpha\beta}=T_{\alpha\beta}\,.
\end{equation}

Let us now focus on the connection field equations which will be obtained from the variation of the Ricci tensor. For a general connection, this variation can be expressed as
\begin{equation}\label{eq:delta R}
 \delta R_{\beta\nu}=\nabla_\lambda\delta\Gamma_{\nu\beta}^\lambda -\nabla_\nu \delta \Gamma_{\lambda\beta}^\lambda+2 \mathcal{T}_{\rho\nu}^\lambda \delta \Gamma_{\lambda\beta}^\rho
\end{equation}
where $\mathcal{T}_{\rho\nu}^\lambda\equiv ( \Gamma_{\rho\nu}^\lambda-\Gamma_{\nu\rho}^\lambda)/2 $ is the torsion tensor. 
Omitting constants for simplicity, the relevant term in (\ref{eq:var_2}) for the connection equations is given by
\begin{equation}
I_\Gamma=\int \d^4x \sqrt{-g} { W^\beta}_\lambda g^{\lambda\nu}\delta R_{\beta\nu} \ .
\end{equation}
Using the formula (\ref{eq:delta R}) in the above equation, integrating by parts and noting that 
\be
\nabla_\mu \sqrt{-g}=\partial_\mu \sqrt{-g}-\Gamma_{\mu\lambda}^\lambda \sqrt{-g}
\ee
for a general connection and the relation for an arbitrary $J^\mu$
\be
\nabla_\mu (\sqrt{-g} J^\mu)=\partial_\mu (\sqrt{-g} J^\mu)-2\mathcal{T}_{\rho\lambda}^\lambda \sqrt{-g} J^\rho
\ee
we finally get (here $W^{\beta\nu}\equiv  { W^\beta}_\lambda g^{\lambda\nu}$)
\begin{eqnarray}
I_{(n)} &=&-\int d^4x\left[  \nabla_\lambda\left( \sqrt{-g}W^{\beta\nu}\right)-\delta_\lambda^\nu \nabla_\rho\left( \sqrt{-g}W^{\beta\rho}\right) \nonumber \right. \\ &+& \left. 2\sqrt{-g}\left(\mathcal{T}_{\lambda\kappa}^\kappa W^{\beta\nu}-\delta^\nu_\lambda \mathcal{T}_{\rho\kappa}^\kappa W^{\beta\rho}+\mathcal{T}_{\lambda\rho}^\nu W^{\beta\rho}\right) \right]\delta \Gamma_{\nu\beta}^\lambda \ ,
\end{eqnarray}
and, therefore, the connection field equations read
\begin{equation}
 \nabla_\lambda\left( \sqrt{-g}W^{\beta\nu}\right)-\delta_\lambda^\nu \nabla_\rho\left( \sqrt{-g}W^{\beta\rho}\right) + 2\sqrt{-g}\left(\mathcal{T}_{\lambda\kappa}^\kappa W^{\beta\nu}-\delta^\nu_\lambda \mathcal{T}_{\rho\kappa}^\kappa W^{\beta\rho}+\mathcal{T}_{\lambda\rho}^\nu W^{\beta\rho}\right) =0
 \label{connectionEq}
\end{equation}
In this derivation we have assumed that the matter sector does not couple to the connection so that it does not contribute to the above equations. From now on and  for simplicity, we shall  restrict ourselves to the case of a symmetric connection, i.e., vanishing torsion tensor. The consistency of this assumption is ensured by the fact that the connection  equations are algebraic in the connection, as we will see below. Setting $\mathcal{T}_{\lambda\rho}^\nu=0$ in (\ref{connectionEq}) we obtain the simplified equations\footnote{Had we set  the torsion to zero in the action, the resulting equations of motion would have been symmetric under the exchange of $\nu$ and $\beta$. The inclusion of the torsion makes the equations not symmetric, which is crucial for the subsequent analysis. See \cite{Olmo:2013lta} for a detailed discussion on this point.}
\begin{equation}
 \left[  \nabla_\lambda\left( \sqrt{-g}W^{\beta\nu}\right)-\delta_\lambda^\nu \nabla_\rho\left( \sqrt{-g}W^{\beta\rho}\right)  \right]=0.
\end{equation}
 Tracing over the indices $\lambda$ and $\nu$, one finds that $\nabla_\rho\left( \sqrt{-g}W^{\beta\rho}\right)=0$, which, plugged back into the equations, implies that the connection equations boil down to 
\begin{equation}
 \nabla_\lambda\left( \sqrt{-g}{ W^\beta}_\rho g^{\rho\nu}\right)=0.
 \label{connectionEq2}
\end{equation}
This equation for the connection can always be solved formally. To do it, we  first note that  the definition of the fundamental matrix $\Mm$ allows to express $\lambda^{-2}R_{\alpha\beta}$  as $\lambda^{-2}R_{\alpha\beta}=g_{\alpha\lambda}{(\hat M^2-\Id)^\lambda}_\beta$. As a result, the metric field equations (\ref{eq:metric_var}) establish an algebraic relation between $\hat M$, the metric and the matter fields, i.e., $\hat M=\hat M(\hat g, \hat T)$. Finding such a relation is the challenging step and, moreover, given the highly non-trivial and non-linear form of the equations, several branches are expected to arise. Furthermore, such a relation implies that $\hat W$ is also an algebraic function of the metric and the matter fields. Therefore, the connection in (\ref{connectionEq2}) only appears linearly and can be solved by algebraic means, i.e., like in General Relativity, it is a non-dynamical object. Solving the resulting equations is, in general, very difficult. This is partially due to the fact that the matrix $\hat{W}\hat{g}^{-1}$ may not be symmetric. In fact, equations (\ref{connectionEq2}) can be decomposed into their symmetric and antisymmetric parts as follows\footnote{Similar equations also appear in the bimetric formulation in GR considered in \cite{BeltranJimenez:2012sz}.}:
\begin{eqnarray}
 \nabla_\lambda\left( \sqrt{-g} g^{\rho(\nu}W^{\beta)}_\rho \right)&=&0, \label{connectionEq2sym}\\
  \nabla_\lambda\left( \sqrt{-g} g^{\rho[\nu}W^{\beta]}_\rho \right)&=&0.
 \label{connectionEq2ant}
\end{eqnarray}
The solution to the symmetric part can be found by simply assuming that there exists a rank-two symmetric tensor $\tilde{g}_{\alpha\beta}$ such that 
\begin{equation}\label{symgW}
\sqrt{-g}{ g^{\rho(\nu}W^{\beta)}}_\rho =\sqrt{-\tilde{g}}\tilde{g}^{\beta\nu}
\end{equation}
which turns (\ref{connectionEq2sym}) into the well-known equation $\nabla_\lambda(\sqrt{-\tilde{g}}\tilde{g}^{\beta\nu})=0$. This equation also appears in the Palatini formulation of General Relativity and establishes the compatibility between the connection and the metric (up to the aforementioned projective symmetry). In our case, it implies that the connection takes the form 
\begin{equation}
\Gamma^\alpha_{\mu\nu}=\frac{\tilde{g}^{\alpha\rho}}{2}\left(\partial_\mu \tilde{g}_{\rho\nu}+\partial_\nu \tilde{g}_{\rho\mu}-\partial_\rho \tilde{g}_{\mu\nu}\right)\,.
\end{equation}
Taking the determinant on both sides of equation (\ref{symgW}) one finds
\begin{equation}
\sqrt{-\tilde{g}}=\frac{(\sqrt{-g})^2}{4}\sqrt{-\det\Big(\Wm\hat{g}^{-1}+\hat{g}^{-1}\Wm^{T}\Big)}
\end{equation}
which leads to 
\begin{equation}
\tilde{g}^{\alpha\beta}=\frac{4}{\sqrt{-g}} \frac{1}{\sqrt{-\det\Big(\Wm\hat{g}^{-1}+\hat{g}^{-1}\Wm^{T}\Big)}} g^{\rho(\nu}W^{\beta)}_\rho .
\end{equation}
In the particular case of symmetric $\Wm\hat{g}^{-1}$ this equation reduces simply to
\begin{equation}\label{eq:hWg}
\tilde{g}^{\alpha\beta}=\frac{1}{\sqrt{\det(\hat W)}}{W^\alpha}_\lambda g^{\lambda\beta} \ \text{ and }  \ \tilde{g}_{\alpha\beta}=\sqrt{\det(\hat W)}g_{\alpha\lambda}{(\hat W^{-1})^\lambda}_\beta  \ .
\end{equation}
Concerning the antisymmetric part of the equations (\ref{connectionEq2ant}), they correspond to the constraint equations that arise from setting the torsion to zero and, indeed, they will determine the consistency of the assumption $\mathcal{T}^\alpha_{\mu\nu}=0$. Notice that having a symmetric matrix $\Wm\gm^{-1}$ trivially fulfills such a constraint. For simplicity, in the following we will assume that such a condition holds.\\

Now we go back to (\ref{eq:metric_var}) and look for a more compact representation of the metric field equations. In matrix notation, this equation can be recast as 
\begin{equation}
\frac{\lambdat^4\lambda^{-2}}{2}
\left[(\hat R \hat {W}) + (\hat R \hat {W})^T\right] ={\LL_G} \hat g+\hat T \ .
\end{equation}
Using the relations (\ref{eq:hWg}), we can write $(\hat R \hat {W})=\det(\hat{W})^{\frac{1}{2}}(\hat {R} \hat {\tilde{g}}^{-1} \hat g)$. Now, since $R_{\alpha\beta}(\Gamma)=R_{\alpha\beta}(\tilde{g})$, it follows that $R_{\alpha\beta}=R_{\beta\alpha}$ as it corresponds to the Ricci tensor of the Levi-Civita connection of a given metric\footnote{We should remember here that having a symmetric connection does not guarantee the  symmetry of the corresponding Ricci tensor. The presence of a non-metricity tensor can give rise to an antisymmetric part of the Ricci tensor, as it happens for instance in Weyl geometries. See for instance \cite{Jimenez:2014rna}.}. We thus see that $(\hat R \hat {W})_{\alpha\beta}=\det(\hat W)^{\frac{1}{2}} {R_\alpha}^\lambda(\tilde{g})  g_{\lambda\beta}$, which turns (\ref{eq:metric_var}) into 
\begin{equation}\label{eq:metric_var_2}
g_{\lambda(\alpha}{R_{\beta)}}^\lambda(\tilde{g})=\frac{\lambda^2}{\lambdat^4 \det(\hat W)^{\frac{1}{2}}}\left(\LL_Gg_{\alpha\beta}+T_{\alpha\beta}\right)  \ .
\end{equation}
When the product $(\hat R \hat {W})$ is symmetric, then the above equation can be further simplified to get
\begin{equation}\label{eq:metric_var_3}
{R_{\mu}}^\nu(\tilde{g})=\frac{\lambda^2}{\lambdat^4 \det(\hat W)^{\frac{1}{2}}}\left(\LL_G {\delta_\mu}^\nu+{T_\mu}^\nu\right)  \ .
\end{equation}
In the remaining of the paper we will illustrate the general properties shown in this section for the simplest case with $\beta_2=\beta_3=\beta_4=0$.


\section{Minimal Born-Infeld extension}\label{min_BI}
Now that we have derived several general properties of the Born-Infeld inspired extensions for gravity considered throughout this paper, we will study the minimal case given by the first two polynomials
\begin{equation}
\mS_{\rm min}=\lambda^2\mpl^2\int \d^4x\sqrt{-g} \Tr\left[\sqrt{\Id+\lambda^{-2}\hat{g}^{-1}\hat{R}}-\Id\right].
\label{eq:gminimal}
\end{equation}
The coefficient $\beta_0$ has been tuned to have Minkowski spacetime as vacuum solution\footnote{As we will show later, even though Minkowski spacetime with vanishing curvature is indeed a vacuum solution, there is another branch that connects with an Einstein space in vacuum due to the non-linear algebraic relation of the curvature with the matter content.}, i.e., to recover Einstein-Hilbert action without a cosmological constant at low curvatures, and  $\beta_1$ has been absorbed into the parameter $\lambdat$. Moreover, as we have also discussed above, to match GR at low curvatures we have identified $\lambdat^4=\mpl^2\lambda^2$.

\subsection{Field equations for the minimal model}
The field equations corresponding to the metric tensor are given by
\be
(M^{-1})^\alpha{}_{(\mu} R_{\nu)\alpha}-\Tr (\Mm-\Id)\lambda^2g_{\mu\nu}=\frac{1}{\mpl^2}T_{\mu\nu}\,.
\ee
If we now take the low curvature limit in these equations with $M^\alpha{}_\beta\simeq \delta^\alpha{}_\beta+\frac12\lambda^{-2}g^{\alpha\sigma}R_{\sigma\beta}$ we obtain
\be
G_{(\mu\nu)}=\frac{1}{\mpl^2}T_{\mu\nu}
\ee
which correspond to the Einstein equations in the Palatini formalism, as expected. However, one should remember that the l.h.s. of these equations correspond to the Einstein tensor for the independent connection. As mentioned above for the general case and will show soon for this particular case, in this limit the connection is nothing but the Levi-Civita connection of the spacetime metric, so we indeed recover GR.

The field equations derived from (\ref{eq:gminimal}) can also be written in matrix notation as
\be
\frac12\left[\Rm\Mm^{-1}+(\Rm \Mm^{-1})^T\right]-\Tr(\Mm-\Id)\lambda^2\gm=\frac{1}{\mpl^2}\Tm.
\ee
Then we can use that $\Rm=\lambda^2\gm(\Mm^2-\Id)$ from the definition of $\Mm$ to express the curvature in terms of the matrix $\Mm$ and obtain the equation
\be
\frac12\left[\gm(\Mm-\Mm^{-1})+(\Mm-\Mm^{-1})^T\gm\right]-\Tr(\Mm-\Id)\gm=\frac{1}{\lambda^2\mpl^2}\Tm.
\label{eq:M}
\ee
This is the equation that allows to obtain the fundamental matrix $\Mm$ by algebraic means in terms of the metric and the matter content, i.e., it gives the function $\Mm(\gm,\Tm)$ mentioned in the previous section. In this special case, it corresponds to a quadratic equation in the elements of $\Mm$, so we expect to find several branches. In general, solving this equation for the matrix $\Mm$ is difficult, so we will consider some simple configurations. In next sections we will assume a perfect fluid form for the matter sector and, therefore, all relevant matrices will be symmetric. In particular, we will have that $\Wm \gm^{-1}$ is symmetric so that the symmetrization of the connection equation in (\ref{connectionEq2}) will not be necessary and its antisymmetric part will be trivially satisfied. For the minimal theory we have that $\Wm=\Mm^{-1}$ and, consequently, the connection is given by the Levi-Civita connection of the effective metric\footnote{Since the effective metric carries a factor depending on the square root of a determinant, one might think that $\tilde{g}_{\mu\nu}$ is a tensorial density. However, since $\Mm$ is a $(1,1)$-rank  tensor, its determinant is not really a tensorial density, but a pure scalar. }
\be\label{effective_metric}
\tilde{g}^{\mu\nu}=\sqrt{\det \Mm}g^{\alpha\mu}(\Mm^{-1})^{\nu}{}_\alpha.
\ee
Notice that, since $\Mm$ is a positive-definite matrix, we have that ${\rm sign} (\tilde{g})={\rm sign}(g)$. Interestingly, for both metrics to be conformally related we need to have $\Mm^{-1}\propto\Id$ that only happens in vacuum as we will show below. Moreover, in such a case, the conformal factor is constant.
 
We have not yet exhausted all the relations that we have. Starting from the definition $q_{\mu\nu}=g_{\mu\nu}+\lambda^{-2}R_{\mu\nu}(\Gamma)$ we can write $\lambda^{-2}R_{\mu\nu}(\Gamma)=q_{\mu\nu}-g_{\mu\nu}$. Since $R_{\mu\nu}(\Gamma)=R_{\mu\nu}(\tilde{g})$, we raise one index of this Ricci tensor with $\tilde{g}^{\mu\nu}$ as follows:
\begin{equation}
\lambda^{-2}{R^\mu}_{\nu}(\tilde{g})=\frac{\sqrt{\det(\hat M)}}{\lambda^2\mpl^2}\left[{(\hat M^{-1} \hat g^{-1} \hat q)^\mu}_{\nu}-{(\hat M^{-1} \hat g^{-1} \hat g)^\mu}_{\nu}\right]=\frac{\sqrt{\det(\hat M)}}{\lambda^2\mpl^2}\left[{(\hat M)^\mu}_{\nu}-{(\hat M^{-1})^\mu}_{\nu}\right] \ ,
\end{equation}
where we used that $ \hat g^{-1} \hat q=\hat M ^2$. Using now the equation of motion for the metric (\ref{eq:M}), we get
\begin{equation}\label{eq:Rmn}
{R^\mu}_{\nu}(\tilde{g})=\sqrt{\det(\hat M)}\mpl^{-2}\left[\LL_{BI_1}{\delta^\mu}_{\nu}+{T^\mu}_{\nu}\right] \ ,
\end{equation}
where $\LL_{BI_1}$ is the gravity Lagrangian of this theory, $\LL_{BI_1}\equiv \lambda^2\mpl^2\Tr\left[\hat{M}-\Id\right]$. Note that the above expression relating the Ricci tensor of the connection with the matter sources is absolutely general, i.e., it is valid for any Palatini theory of gravity (with just small variations). All the elements on the right-hand side of this equation are functions of the matter sources. Applications to particular scenarios are now possible, but before that let us  end this section by signaling how general bounds on the energy density can arise. If we take the trace of the eq. (\ref{eq:M}) with respect to $g^{\mu\nu}$ we find
\be
\Tr\left(\Mm^{-1}+3\Mm\right)=16\left(1-\frac{g^{\alpha\beta} T_{\alpha\beta}}{16\lambdat^4}\right)
\ee
where we have used that $\lambda^2\mpl^2=\lambdat^4$. This allows to impose a constraint on the energy-momentum tensor. Since the matrix $\Mm$ is positive-definite, the l.h.s. of the above equation is positive and, thus, we necessarily have that
\be
\frac{g^{\alpha\beta} T_{\alpha\beta}}{16\lambdat^4}=\frac{3p-\rho}{16\lambdat^4}\leq1.
\ee
For a radiation-like fluid, this constraint is trivially satisfied. For a dust component, the condition $\lambdat^4<0$ also allows to trivially fulfill the constraint, whereas if $\lambdat^4>0$ there is an upper bound for the allowed energy-densities given by $16\lambdat^4$. However, it is important to notice that although this is a necessary constraint, it does not need to give the physical upper bound for $\rho$ and, in fact, there are more stringent constraints, as we will see in the following. On the other hand, the constraint will depend on the type of matter and it is not guaranteed the existence of an upper bound for $\rho$ or $p$ for a general matter component. The generality of these results and the specific bounds for different fluids will be shown in more detail in the following sections.

\subsection{Einstein space solutions}
Let us start by considering an Einstein space ansatz with $R_{\mu\nu}(\Gamma)=\kappa g_{\mu\nu}$. For purely Einstein spaces $\kappa$ is a constant, but here we will relax this assumption and let $\kappa$ be an arbitrary function of space and time for the moment and show that it needs to be constant. The form of the Ricci tensor implies that $M^\alpha{}_\beta=m^2\delta^\alpha{}_\beta$ with $m^2\equiv\sqrt{1+\lambda^{-2}\kappa}$. Then, from (\ref{eq:M}) we obtain
\be
\label{ESminimal}
\left(4-3m^2-\frac{1}{m^2}\right)g_{\mu\nu}=\frac{1}{\lambdat^4}T_{\mu\nu}.
\ee
If we take the covariant derivative of these equations with respect to the Levi-Civita connection of $g_{\mu\nu}$, the conservation of the energy momentum tensor of matter\footnote{Remember that we are assuming matter fields minimally coupled to the space-time metric $g_{\mu\nu}$ from where it follows the conservation of the energy momentum tensor.} implies that $m^2$ (and therefore $\kappa$) must be a constant. Hence, only a cosmological constant-like fluid with $T_{\mu\nu}=-\rho_{\Lambda}g_{\mu\nu}$ with $\rho_\Lambda$ constant in space and time can support Einstein space solutions. By taking the trace of equation (\ref{ESminimal}) with respect to $g^{\mu\nu}$ we obtain an equation for $m^2$ in terms of the trace of the energy-momentum tensor
\be
4-3m^2-\frac{1}{m^2}+\frac{\rho_\Lambda}{\lambdat^4}=0
\ee
whose solutions are
\be\label{solutions_M_ES}
m^2=\frac{4+\tilde{\rho}_\Lambda\pm\sqrt{4+\tilde{\rho}_\Lambda(8+\tilde{\rho}_\Lambda)}}{6}
\ee
where $\tilde{\rho}_\Lambda$ stands for $\tilde{\rho}_\Lambda=\rho_\Lambda/\lambdat^4$. Now that we know the solution for the fundamental matrix $\Mm$ in terms of the matter content and the metric tensor, we can solve the connection equations. Since $\Mm$ is symmetric (diagonal in fact), we know that there is a solution given by the Levi-Civita connection of the metric defined by (\ref{effective_metric}) and the antisymmetric part of the connection equations is trivially satisfied. Such effective metric is given in this case by
\be
\tilde{g}_{\mu\nu}=\frac{1}{m^2}g_{\mu\nu}\,,
\ee
so that both metrics are simply related by means of a constant conformal transformation.  Then, we have that $R_{\mu\nu}(\tilde{g})=m^2\kappa  \tilde{g}_{\mu\nu}$. Since both metrics are  related by the above constant conformal transformation, we also have that $R_{\mu\nu}(g)=R_{\mu\nu}(\tilde{g})$.

\begin{figure}[h!]
\begin{center} 
\includegraphics[width=16cm]{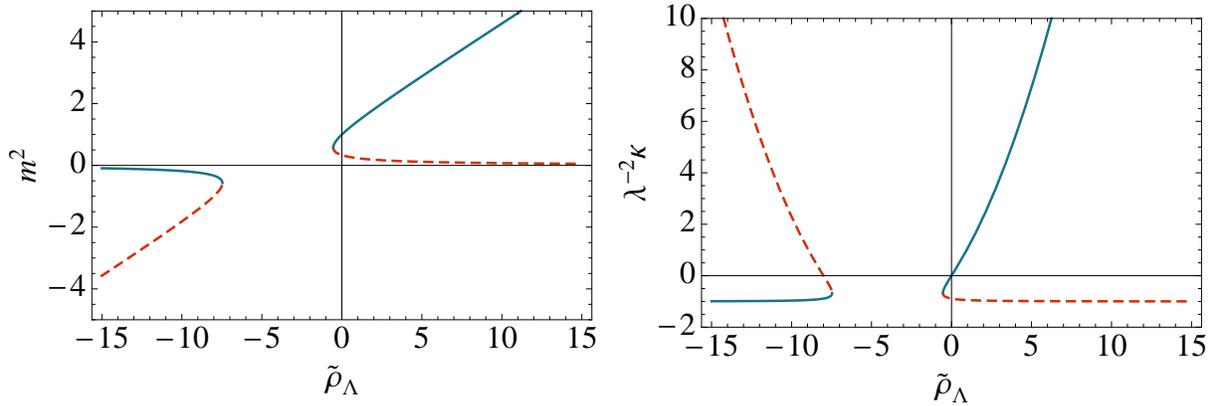}
\caption{In this plot we show the dependence of $m^2$ and $\kappa$ as a function of $\tilde{\rho}_\Lambda$. We see the two branches discussed in the main text: Branch I that is connected to vanishing Ricci tensor in vacuum (blue-dashed) and Branch II with dS/AdS (red-solid). Since $\Mm$ is a positive definite matrix, only those values of $\rhot_\Lambda$ for which $m^2$ is positive are physical. For a very large cosmological constant, the curvature $\kappa$ saturates to the value $-\lambda^2$.}
\label{m2Einstein}
\end{center}
\end{figure}

\subsubsection{Vacuum solutions}
 In vacuum we have $\rho_\Lambda=0$ and the solutions are $m^2=1$ and $m^2=1/3$, i.e., $\kappa=0$ and $\kappa=-\frac{8\lambda^2}{9}$. Whereas the former represents a Ricci-flat space with $R_{\mu\nu}(\Gamma)=0$ (Branch I), the latter gives a non-vanishing Ricci tensor $R_{\mu\nu}(\Gamma)=-\frac{8\lambda^2}{9}g_{\mu\nu}$ and constant curvature $R=g^{\mu\nu}R_{\mu\nu}(\Gamma)=-\frac{32\lambda^2}{9}$ (Branch II). Although we have obtained these solutions by imposing the Einstein space Ansatz from the beginning, they correspond to the general solutions in vacuum with $T_{\mu\nu}=0$. This is easy to see because in vacuum the fundamental matrix $\Mm$ must be proportional to $\delta^\alpha{}_\beta$ and the Ricci tensor must take the Einstein space form. It is interesting to note that Branch II corresponds to a de Sitter or anti-de Sitter space for $\lambda^2$ negative  and positive respectively without having a cosmological constant. In the next subsection we study the solutions in the presence of a cosmological constant.
 
 \subsubsection{Cosmological constant solutions}
When $\rho_\Lambda\neq0$, the solutions are given by (\ref{solutions_M_ES}) and are shown in Figure \ref{m2Einstein}. As we can see in that Figure, we obtain two branches which are connected to the Ricci-flat (Branch I) and the dS/AdS (Branch II) solutions in vacuum. Only Branch I gives GR in the limit of low curvatures (or $\lambda\rightarrow\infty$). For the solutions in both branches to be physically viable, we need to have $m^2$ real and positive to guarantee that $\Mm$ is positive definite. Both conditions are fulfilled for $\tilde{\rho}_\Lambda\geq2(\sqrt{3}-2)$ (see Figure \ref{m2Einstein}). Notice that the constraint translates into a bound for a negative (positive) cosmological constant for $\lambda^2$ positive (negative).

In this simple case we already encounter the first saturation effect typical of Born-Infeld-like theories. In the presence of a very large cosmological constant with $\tilde{\rho}_\Lambda\gg1$ we have that $m^2\rightarrow\infty$ for the Branch I that is connected with vanishing Ricci tensor in vacuum, whereas $m^2\rightarrow0$ in the Branch II that connects to dS/AdS. The former case generates an infinite curvature as the cosmological constant value goes to infinity, but in the latter case we have that $\lambda^{-2}\kappa\rightarrow-1$ which means that only a curvature $R(\Gamma)=4\kappa=-4\lambda^{-2}$ is actually generated despite having an arbitrarily large cosmological constant, i.e., only an effective cosmological constant $\Lambda_{\rm eff}=-\lambda^{-2}$ is active. Notice that this saturation is reached very quickly as we increase $\rhot_\Lambda$ and, in fact, it is essentially insensitive to it. Thus, no matter what the value of the cosmological constant is (provided it satisfies the aforementioned bound), in the Branch II the solution is always a de Sitter or an anti-de Sitter space whose curvature is entirely determined by $-\lambda^2$. If $\lambda^2$ is positive (negative), we can have a very large and positive (negative) cosmological constant, but the curvature would correspond to an anti-de Sitter (de Sitter) space with $\Lambda_{\rm eff}\simeq -\lambda^2$. Unfortunately, this mechanism does not look very promising to help with the cosmological constant problem because it occurs in the branch that does not recover GR for the low curvature limit. Hence, the cosmological evolution in such a branch should be compatible with the standard thermal history of the universe, which seems unlikely precisely because of the non-recovery of GR. Moreover, as we will see, this saturation effect and the dS/AdS solutions that we have described here in the presence of a cosmological constant will remain valid for a broad set of equations of state.

\subsection{Perfect fluid solutions}

\begin{figure}[h!]
\begin{center} 
\includegraphics[width=16cm]{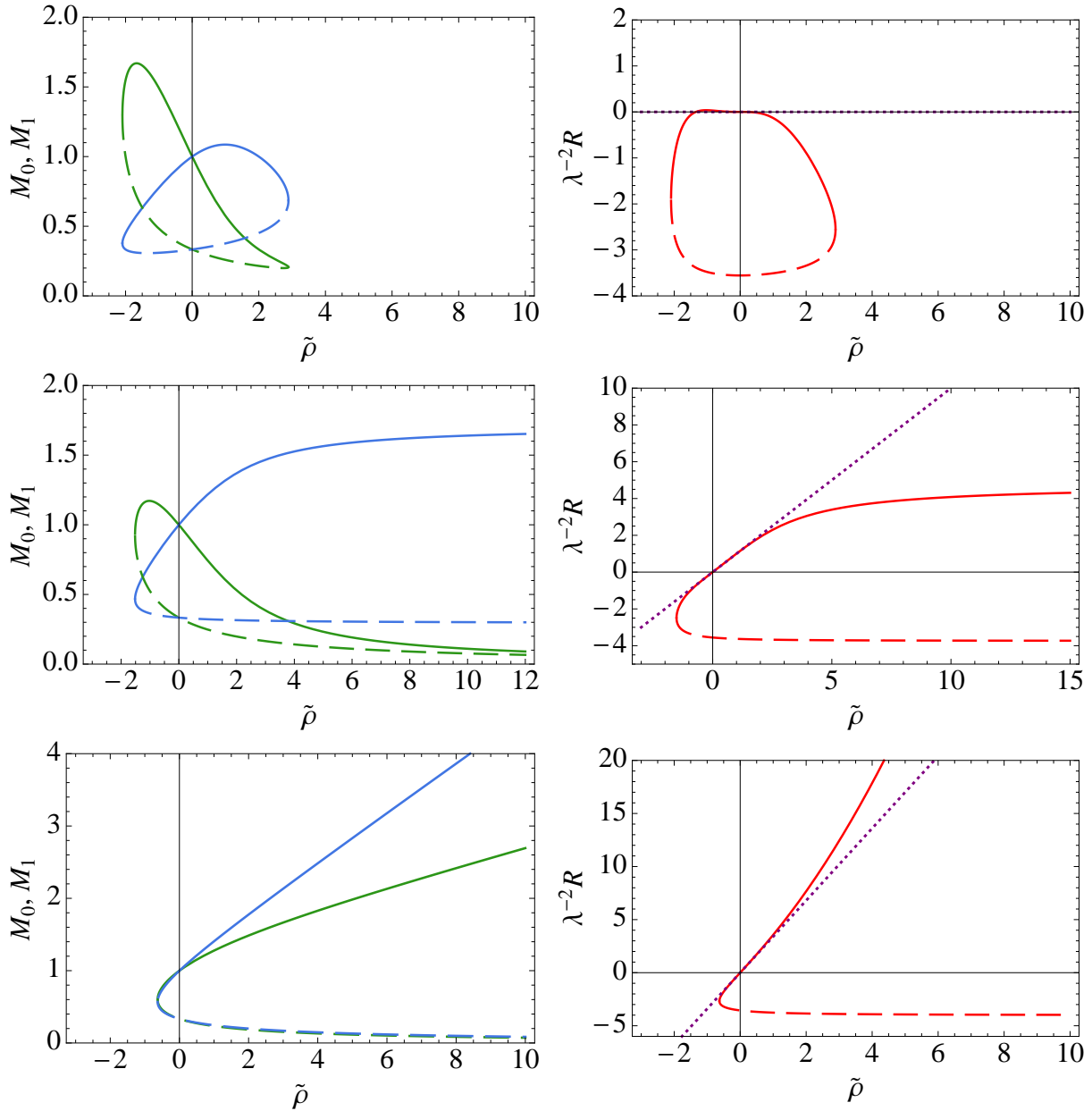}
\caption{This figure illustrates the two branches of solutions (Branch I and II in solid and dashed lines respectively) for $M_0$ (green) and $M_1$ (blue) in the left panels and the scalar curvature $R$ in the right panels  as a function of the energy density of the perfect fluid (normalized with $\lambda^2\mpl^2$). We have considered fluids with equations of state $p=1/3 \rho$ (upper panels), $p=0$ (middle panels) and $p=-0.8\rho$ (lower panels). Here we only show the regions for which the solutions are physical, i.e., with both $M_0$ and $M_1$ real positive quantities. We also plot the GR solutions in dotted-purple line for comparison and we can clearly see when the Born-Infeld model deviates from GR in Branch I. }
\label{perfectFluid}
\end{center}
\end{figure}

In this section we will pay our attention to the case of a perfect fluid with energy-momentum tensor described by
\begin{equation}
{T^\mu}_{\nu}=\left(\begin{array}{cc} -\rho
 & \vec{0} \\
\vec{0} & p \Id_{3\times 3}
\end{array}\right)  \ .
\end{equation} 
Since this matter source is diagonal, the central object of the theory, namely the matrix $\Mm$, will also be assumed to take a diagonal form as follows:
\begin{equation}
{M^\mu}_{\nu}=\left(\begin{array}{cc} M_0
 & \vec{0} \\
\vec{0} & M_1 \Id_{3\times 3}
\end{array}\right)  \ ,
\end{equation} 
where $M_0$ and $M_1$ must be positive to guarantee the positiveness of $\Mm$. So far we have only made the assumption of the perfect fluid form for the matter source and, consequently,  our results will be valid for general inhomogeneous configurations (e.g. spherically symmetric). In next section we will further assume homogeneity in order to study cosmological scenarios. 

The metric field equations written as in (\ref{eq:M}) give 
\begin{eqnarray}
\frac{1}{M_0}+3M_1&=&4+\tilde{\rho}\\
M_0+2M_1+\frac{1}{M_1}&=&4-\tilde{p}
\end{eqnarray}
where we have again defined the dimensionless quantities  $\rhot\equiv\frac{\rho}{\lambda^2\mpl^2}$ and $\tilde{p}\equiv\frac{p}{\lambda^2\mpl^2}$. As we mentioned above, these equations allow to obtain $M_0$ and $M_1$ in terms of the matter content $\rho$ and $p$ by algebraic means. If we isolate $M_1$ from the first equation 
\be
M_1=\frac{1}{3}\left(4+\rhot-\frac{1}{M_0}\right)
\ee
and substitute in the second one we obtain the following equation for $M_0$:
\be
3(4+\rhot) M_0^3-\Big[3\pt(4+\rhot)+2(-5+2\rhot+\rhot^2)\Big]M_0^2-\Big[3\pt+4(1+\rhot)\Big]
M_0+2=0.
\ee
Since this is a cubic equation, there will always be at least one real solution. However, we need to guarantee that such a solution is also positive to fulfill the physical requirements for $\Mm$. In vacuum with $\tilde{\rho}=\tilde{p}=0$ the solutions to the above equations are $M_0^{\rm I}=M_1^{\rm I}=1$, $M_0^{\rm II}=M_1^{\rm II}=1/3$ and $M_0^{\rm III}=-1/2$, $M_1^{\rm III}=2$. Only the Branches I and II can be physical, so we will disregard Branch III from our analysis. As before, Branch I is connected to the Ricci flat solutions in vacuum whereas Branch II is connected to a dS/AdS space, in agreement with our findings in the previous section for the vacuum solutions.

For non-vacuum solutions, the equations for $M_0$ and $M_1$ can still be solved analytically, although their explicit expressions are lengthy and not too illuminating. Once the solutions for $M_0$ and $M_1$ are found, we can easily compute the scalar curvature by using the definition of the fundamental matrix $\Mm$ to obtain
\be
R(\Gamma)=g^{\mu\nu}R_{\mu\nu}(\Gamma)=\lambda^2\left(M_0^2+3M_1^2-4 \right).
\ee
In the limit $\rhot\rightarrow0$ we obtain:
\begin{eqnarray}
R^{\rm I}&=&\frac{\rho-3p}{\mpl^2}+\Od\Big(\frac{\rho^2}{\lambda^4\mpl^4}\Big)\\
R^{\rm II}&=&-\frac{32}{9}\lambda^2-\frac{1}{9\mpl}(\rho-3p)+\Od\Big(\frac{\rho^2}{\lambda^4\mpl^4}\Big)
\end{eqnarray}
confirming that Branch I connects with the GR result at low energy densities whereas Branch II is connected with the dS/AdS branch. 

In Figure \ref{perfectFluid} we plot the solutions for the cases of a radiation fluid with $p=1/3\rho$, pressure-less dust ($p=0$) and a fluid with equation of state $w=-0.8$. We see that for the radiation component there is un upper bound on the allowed energy densities irrespectively of the sign of $\lambda^2$ for both branches. This also implies an upper bound for the possible curvatures. This feature can be relevant for early universe cosmology because as we go back in time, the energy density of the relativistic degrees of freedom (which in fact become more numerous as the temperature of the universe increases) grows, but such a growth is limited by the value of $\lambda^2$ in the minimal Born-Infeld theory. This behaviour is found for all fluids with\footnote{We restrict ourselves to fluids with equations of state parameters such that $\vert w\vert\leq1$.} $0<w<1$.

For the dust component, only when $\lambda^2<0$ there is un upper bound for the energy density. However, if $\lambda^2>0$ the energy density of the dust fluid can be arbitrarily large, although the scalar curvature saturates to a value of order $\lambda^2$ in both physical branches. In fact, in the limit of high densities $\rhot\rightarrow\infty$, the curvature is given by $R_{\rm I,II}=(\frac12\pm3\sqrt{2})\lambda^2+\Od(\rhot^{-1})$. These results could be applied to collapsing objects and the formation of black holes singularities. For negative $\lambda^2$, the Born-Infeld structure of the action prevents the energy density to grow above $\sim\lambda^2\mpl^2$, whereas if $\lambda^2>0$, the energy density can be arbitrarily large but this does not correspond to a curvature singularity, which saturates to a value $\sim\lambda^2$.

In the case of $w=-0.8$, Branch II behaviour is essentially the same as  the one explained for the case of a dust fluid. In fact this is generic for fluids with $w<0$, showing that the solution in Branch II is quite insensitive to the particular fluid that fills the spacetime. In Branch I we see that the curvature grows without bound so that the saturation effect is not present. As a matter of fact, in the high density limit we find $R^{\rm I}\propto \rho^2/(\lambda^2\mpl^4)$, which represents a growth faster than in GR where $R_{\rm GR}=(1-3w)\rho/\mpl^2$ for all energy densities. These general features are actually found for fluids with $w<0$.

\subsection{Cosmological evolution}\label{cosmology_min_BI}

In this section we will analyze the cosmological evolution for the minimal extension of the Born-Infeld inspired theory under study. For that we will take the results from the previous section and consider the further assumption of homogeneous and isotropic configurations. Thus, we will take the FLRW Ansatz for our metric
\begin{eqnarray}
\d s^2=-N(t)^2\d t^2+a(t)^2 \delta_{ij} \d x^{i}\d x^j .
\end{eqnarray}
As explained above, the connection can be identified with the Levi-Civita connection of an effective metric given in (\ref{effective_metric}). For the FLRW Ansatz, the effective metric takes the following form:
\be
\d \tilde{s}^2=-N^2(M_0 M_1^{-3})^{1/2}\d t^2+\frac{a(t)^2}{\sqrt{M_0 M_1}}\delta_{ij}\d x^i\d x^j
\label{eq:effectiveFLRW}
\ee
which is again an FLRW metric but with modified lapse function and scale factor:
\be
\tilde{N}^2\equiv  N^2\sqrt{M_0M_1^{-3}}\, ,\quad    \tilde{a}(t)^2\equiv\frac{a(t)^2}{\sqrt{M_0 M_1}}.
\ee
Since we are considering a perfect fluid as matter source, the solutions for $M_0$ and $M_1$ that we found in the previous section will be the solutions for the present case. Knowing the matrix $\hat M$ in terms of the matter density and pressure will allow us to compute the Hubble function $H\equiv\frac{\dot{a}}{a}$ using the field equations (\ref{eq:Rmn}). For that we first express the Einstein tensor of $\tilde{g}_{\mu\nu}$ in terms of the fundamental matrix $\Mm$
\begin{eqnarray}\label{eq:Gmn}
\hat{G}(\tilde{g})&\equiv&\Rm(\tilde{g})-\frac12\,\hat{\tilde{g}}\, \Tr( \hat{\tilde{g}}^{-1} \Rm)\nonumber \\
&=& \lambda^{2}\hat{g}\left[ \Big(\Mm^2-\Id\Big) -\frac12\Mm \Tr\Big(\Mm-\Mm^{-1}\Big)\right]
\end{eqnarray}
where we have used that $\Rm=\lambda^2\hat{g}\Big(\Mm^2-\Id\Big)$, $\hat{\tilde{g}}=\hat{g}\Mm / \sqrt{\det(M)}$ and $\hat{\tilde{g}}^{-1}= \sqrt{\det(M)}\Mm^{-1}\hat{g}^{-1}$.\\ 
The Hubble function can be determined from the zero-zero component of this Einstein tensor
\begin{equation}\label{compG00}
G_{00}(\tilde{g})=3\left(\frac{\dot{\tilde{a}}}{\tilde{a}}\right)^2=3\left( H+\frac{\dot{\mathcal{A}}}{\mathcal{A}}\right)^2=H^2\Big[ 1-3(\rho+p)\partial_\rho \ln(\mathcal{A}) \Big]^2
\end{equation}
with $\mathcal{A}^2=(M_0M_1)^{-1/2}$, i.e. $\tilde{a}^2=a^2\mathcal{A}^2$, and we have used the fact that $\dot{\mathcal{A}}/\mathcal{A}=\partial_\rho \ln(\mathcal{A})\dot{\rho}$ together with the continuity equation for the fluid $\dot{\rho}=-3H(\rho+p)$. Now plugging the expression (\ref{compG00}) for $G_{00}$ into the zero-zero component of the equation (\ref{eq:Gmn}) enables us to write the Hubble function in terms of the matter field
\begin{eqnarray}
\lambda^{-2}H^2=\frac{1-M_0^2+3M_0 M_1-\frac{3M_0}{M_1}}{6\left(1-3(\rho+p)\partial_\rho \ln(\mathcal{A}) \right)^2} 
\end{eqnarray}
where we have fixed the lapse function to $N(t)=1$. We must remember here that  $M_0$ and $M_1$ are functions of the energy density (and equation of state parameter) of the matter component that can be taken from the results obtained in the previous subsection. Here we will focus on Branch I of solutions because it is the one connected with GR at low energy densities. Moreover, we will also assume $\lambda^2>0$. 

\begin{figure}[ht!]
\begin{center} 
\includegraphics[width=12cm]{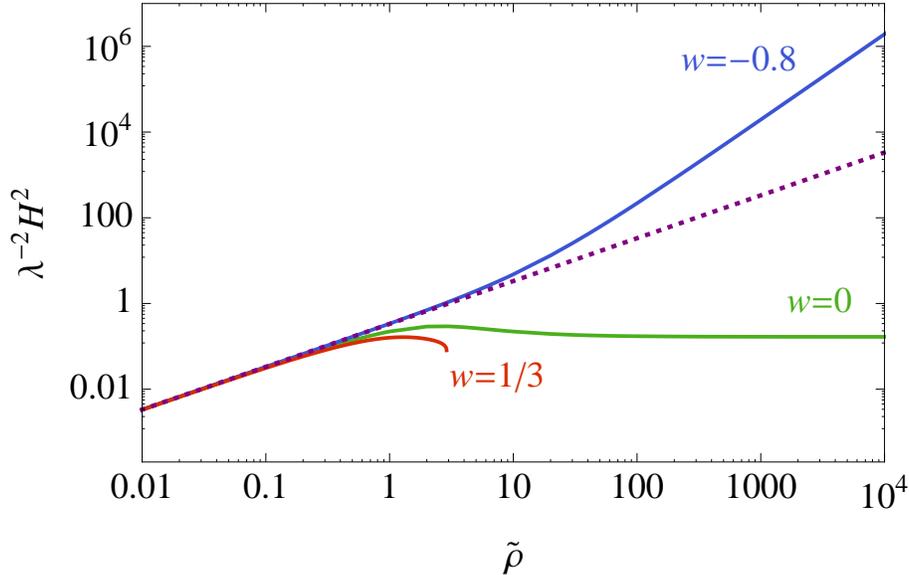}
\caption{In this plot we show the dependence of the Hubble expansion rate as a function of $\rhot$ for three fluids with different equations of state to illustrate the three regimes discussed in the main text. We also plot the solution for GR (purple) for comparison. We see how in the case of a radiation component there is a bound for the energy density. In this plot we have focused on Branch I (connected with the GR limit at low densities) and we have taken $\lambda^2>0$.}
\label{H2I}
\end{center}
\end{figure}

In Figure \ref{H2I} we show the dependence of the Hubble function on the energy density of the fluid for different equation of state parameters, namely $w=1/3$ (radiation), $w=0$ (dust) and $w=-0.8$. For the case of radiation we encounter again the upper bound on the allowed energy density and, in addition, this is the typical behaviour for fluids with $w>0$. As commented already above, this is an interesting feature for early universe cosmology since such an upper bound for the energy density might actually prevent the formation of Big Bang singularities. For dust we see again the saturation effect that we already discussed in the solutions for perfect fluids that causes $H^2$ to go to a constant for very high energy densities. Moreover, this behaviour is shared by all fluids with equation of state parameter $-2/3<w\leq0$. It is important to note that the Hubble function $H^2$ is related to the curvature of the Levi-Civita connection of the spacetime metric and, therefore, is in general different from the total connection curvature $R(\Gamma)$. Thus, although the total curvature only goes to a constant at high energy densities for a dust component, the Hubble expansion rate saturates for equations of state $0\leq w<-2/3$.  Finally, for fluids with $-1<w<-2/3$, we have that $H^2\propto \rho^2/(\lambda^2\mpl^4)$ for high densities so that it grows faster than in GR.

The fact that the Hubble function is constant for very high energy densities can represent an interesting mechanism to generate a de Sitter inflationary era in the presence of a matter component crucially different from an effective cosmological constant. Thus, one could for instance develop such an inflationary epoch in a universe filled with a dust component. Work is in progress in this direction and will be presented elsewhere.


\section{Summary and discussion}
In this paper we have proposed an extension of the modified gravity theories originally inspired by the Born-Infeld electromagnetism as a possible mechanism to regularize curvature singularities. Our extension to such theories are motivated by the massive gravity interactions where the potential is constructed in terms of the elementary symmetric polynomials of a fundamental object given by the square root of a very specific matrix. The original Born-Infeld gravity action can be expressed as the determinant of the matrix $\Omegam=\sqrt{\hat{g}^{-1}\hat{q}}$ with $\hat{q}\equiv \hat{g}+\lambda^{-2}\hat{R}$ so that it represents the forth order polynomial associated to the matrix $\Omegam$. Therefore, it is a natural generalization of the Born-Infeld gravity to include all of the elementary symmetric polynomials of such a matrix. In this way, we adapt lessons from an IR modification of the gravitational interaction (massive gravity) to the case of Born-Infeld inspired gravity theories, which modify gravity at high curvatures. 

After introducing the extended version of the Born-Infeld gravity, we have computed the corresponding field equations. We treat the theory in the Palatini formalism so that the connection is treated as an independent object in the theory. As usual in the Palatini formalism, the connection can actually be algebraically solved so that it represents an auxiliary field which does not add new degrees of freedom with respect to the metric formalism. We managed to solve the connection as the Levi-Civita connection of an effective metric that is given in terms of the space-time metric tensor and the matter content by means of an algebraic equation. This equation is in general highly non-trivial and leads to the existence of several branches of solutions. We show that one of these branches is always continuously connected with General Relativity at low curvatures.

Once we have presented the aforementioned general results for the full theory, we have studied in more detail a minimal version of it consisting of only the first two polynomials with the condition that Ricci-flat solutions exist in vacuum. For this model we have analysed Einstein space solutions and show that they can only be supported by a cosmological constant. In vacuum (vanishing cosmological constant) we have obtained the usual Ricci-flat solution (as imposed from the construction of the theory), but also an additional dS/AdS solution with a curvature of order $\lambda^2$ corresponding to the branch that is not connected with the GR solution. Additionally, for an arbitrary cosmological constant we found that this dS/AdS solution remains the same irrespectively of the value of the cosmological constant. In fact, we have shown that this solution is quite general for arbitrary fluids with equation of state $w\le0$.

We then have looked at solutions with a perfect fluid as matter source, for which it was also possible to solve the connection equations. We found that for a radiation fluid (and in general for fluids with $0<w\leq1$) there is un upper bound on the possible energy densities and this could have an important effect for early universe cosmology. For dust we obtained a saturation effect characteristic of Born-Infeld-like actions where the curvature remained constant at very high energy densities. This could have interesting consequences for the gravitational collapse and the formation of black hole singularities. For equation of state parameters smaller than zero, we found that the curvature grows as $\rho^2$ instead of going like $\rho$ as in GR. Finally, we have looked at some cosmological scenarios with homogeneity and isotropy. In that case, we could fully solve the field equations and find the effective metric that generates the space-time connection. Moreover, we also obtained an expression for the Hubble expansion rate in terms of the energy density of the fluid. For a universe filled with radiation we found once again the upper bound for the possible energy densities. Interestingly, for equation of state parameters $-2/3<w\leq0$, the Hubble function remains constant at very high densities so that it would be possible to have a (quasi) de Sitter inflationary phase in a universe filled with more general fluids than in GR (e.g., in a dust-dominated universe). This is a consequence of the Born-Infeld structure of the action and its viability as a successful inflationary phase will be studied in detail elsewhere.

We have focused on the minimal version of the extended version of the Born-Infeld inspired theories as the simplest case, but the general theory with all the polynomials will be interesting to study. In particular, more physically viable branches are expected to arise due to the higher degree of the field equations. One motivation to study Born-Infeld-like theories within the context of modified gravity is the potential regularization of the singularities. Here we have shown that, although the regularization is not a general feature of these theories, for some interesting cases like Big-Bang or black hole singularities, such a regularization might actually take place. However, these issues should be studied more carefully before drawing any final conclusion on it.



\acknowledgments
J.B.J. and L.H. wish to acknowledge the Departamento de F\'isica Te\'orica and IFIC at Universidad de Valencia-CSIC  for hospitality and support at the initial stage of this work. 
J.B.J. is supported by the Wallonia-Brussels Federation grant ARC No. 11/15-040 and also acknowledges financial support from MINECO (Spain) projects FIS2011-23000 and Consolider-Ingenio MULTIDARK CSD2009-00064.
L.H. is supported by the Swiss National Science Foundation. G.J.O. is supported by the Spanish Grant No. FIS2011-29813-C02-02, the Consolider Program CPANPHY-1205388, the JAE-doc program and Grant No. i-LINK0780 of the Spanish Research Council (CSIC), and by CNPq (Brazilian agency) through Project No. 301137/2014-5.


\bibliographystyle{JHEPmodplain}
\bibliography{references}

\end{document}